\documentclass[letterpaper, 10 pt, conference]{ieeeconf}  

\IEEEoverridecommandlockouts                              
\newcommand{\R}{ \mathbb{R} }

\usepackage{graphics} 
\usepackage{epsfig} 
\usepackage{mathptmx} 
\usepackage{times} 
\usepackage{amsmath} 
\usepackage{amssymb}  
\usepackage{tikz}
\usepackage{multirow}
\newtheorem{theorem}{\textbf{Theorem}}[section]

\newtheorem{remark}{\textbf{Remark}}
\title{\LARGE \bf
A Parallel in Time Algorithm Based on ParaExp for Optimal Control Problems 
}

\author{Felix Kwok 
and Djahou N. Tognon
\thanks{F.~Kwok, Université Laval, Département de mathématiques et de statistique, Canada 
        {\tt\small felix.kwok@mat.ulaval.ca}}%
\thanks{D.~N.~Tognon, INRIA Paris : ANGE Project-Team,
and Sorbonne Université: Laboratoire Jacques-Louis Lions (LJLL), France 
        {\tt\small djahou-norbert.tognon@inria.fr}}%
}

\begin{document}
\maketitle
\thispagestyle{plain}
\pagestyle{plain}

\begin{abstract}
We propose a new parallel-in-time algorithm for solving optimal control problems constrained by discretized partial differential equations.  Our approach, which is based on a deeper understanding of ParaExp, considers an overlapping time-domain decomposition in which we combine the solution of homogeneous problems using exponential propagation with the local solutions of inhomogeneous problems. The algorithm yields a linear system whose matrix-vector product can be fully performed in parallel. We then propose a preconditioner to speed up the convergence of GMRES in the special cases of the heat and wave equations. Numerical experiments are provided to illustrate the efficiency of our preconditioners.
\end{abstract}

\section{Introduction}
The efficient solution of optimal control problems constrained by time-dependent partial differential equations (PDEs) has received much interest in recent decades, given its importance in applications and the 
increasing availability of powerful modern computing clusters.
The main computational challenges lie in the large amounts of data that must be processed, the intensive computation required to simulate the underlying PDEs, and the optimization procedure needed to determine the best
control function. Parallel numerical algorithms, specifically those of the domain decomposition type, are 
well suited for tackling these huge problems by subdividing them into manageable
chunks that can be handled simultaneously by many processors. Parallelism also has the obvious advantage of
reducing the wall-clock time required to find the optimal solution, which is especially important 
for real-time applications.

There exists a rich literature on spatial domain decomposition
methods for elliptic and initial value problems, see \cite{toselli2005domain,gander2007optimized,hoang2013space,li2015convergence} and the references therein.
There is also a growing literature on parallel-in-time methods, in which multiple processors solve the initial value problem on different parts of the time interval simultaneously;
see \cite{gander201550,ong2020applications} for a survey of these methods. Time parallelization can also be applied to optimal control problems, either
by decomposing the optimization problem directly, see \cite{heinkenschloss2005time,maday2007monotonic,barker2015domain,gander2016schwarz}, or by using a parallel-in-time method to solve
the forward and adjoint problems required by gradient computations \cite{ulbrich2007generalized}.

In this paper, we focus on the efficient solution of linear-quadratic optimal control problems when a cheap exponential integrator is available, i.e., when one is able to evaluate $\exp(t{\cal L})$ quickly for a matrix ${\cal L}$
arising from spatial discretization. For initial value problems, the authors of \cite{doi:10.1137/110856137} have used
such integrators to construct a parallel-in-time method known as ParaExp. In practice,
many different methods can be used to evaluate $\exp(t{\cal L})$, and the choice depends on the particular
properties of the operator ${\cal L}$. In this paper, we show how similar ideas can be used to solve optimal
control problems; we will focus on the particular cases of the heat and wave equations, where specific approximations are used to construct the preconditioner for solving for the optimal adjoint state. Numerical experiments are provided to illustrate the behavior of these preconditioners.


\section{Linear optimal control problem}
We consider the optimal control problem constrained by a system of ordinary differential equations (ODEs)
\begin{gather*}
\min_{\nu} \frac12\left\|y(T)-y_{tg}\right\|^2+\frac{\alpha}{2}\int_{0}^T\left\|\nu(t)\right\|^2dt,\\
\textrm{subject to}\qquad \dot y(t)={\cal L}y(t)+\nu(t), \; y(0)=y_{in}, \, t\in (0,T),\nonumber
\end{gather*}
where the constraint arises from a semi-discretization in space of a time-dependent PDE.
Here, $\alpha$ is a regularization parameter and $y_{in}$, $y_{tg}$ are the initial and target states respectively. The state function $y$ and the control $\nu$ are mappings from $(0, T)$ to $\R^r$, $r$ a positive integer. The operator ${\cal L} \in \R^{r\times r}$ is independent of the time variable.  Using the Lagrange multiplier approach, 
one can characterize the optimal trajectory $y(t)$ via the forward and
adjoint problems  
\begin{equation*}
\begin{cases}
\dot y={\cal L}y+\nu\quad \textrm{on}\;\, (0,T),\\
y(0)=y_{in},
\end{cases}
\qquad \begin{cases}
\dot\lambda=-{\cal L}^T\lambda\quad \textrm{on}\;\, (0,T)\\
\lambda(T)=y(T)-y_{tg},
\end{cases}
\end{equation*}
where the control $\nu$ and adjoint state $\lambda$ are related by the algebraic equation $\lambda(t)=\alpha\nu(t)$ for all $t \in (0,T)$. Eliminating $\nu$, the above system can also be written as 
\begin{equation}\label{eq:systemoptimality}
\dot y={\cal L}y-\frac1\alpha\lambda, \quad \dot\lambda=-{\cal L}^T\lambda \quad \textrm{on}\;\, (0,T) 
\end{equation}
with the initial and final condition $y(0)=y_{in}$ and $\lambda(T)=y(T)-y_{tg}$ respectively. 
\section{Parallel-in-time algorithm} 

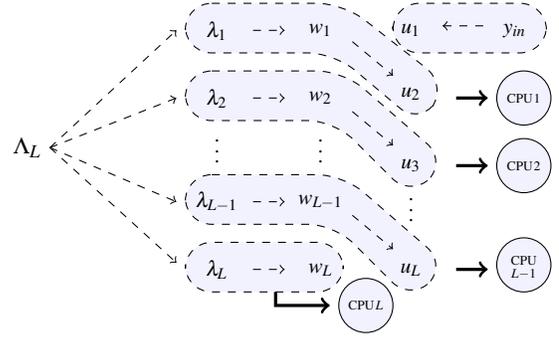
\begin{figure}[h!]
        \centering
	\begin{tikzpicture}[scale=0.6]

		\draw[fill=blue!5] (10,3.1) circle(0.6);
		\draw (10,3.1) node{\tiny ${\mathrm{CPU}\, 1}$};
		
		\draw[fill=blue!5] (10,1.6) circle(0.6);
		\draw (10,1.6) node{\tiny ${\mathrm{CPU}\,2}$};
		\draw[fill=blue!5] (10.,-0.60) circle(0.6);
		\draw (10,-.50) node{\tiny $\mathrm{CPU}$};
		\draw (10,-.8) node{\tiny $L\!-\!1$};
		\draw[fill=blue!5] (6.5,-1.5) circle(0.6);
		\draw (6.5,-1.5) node{\tiny ${\mathrm{CPU}\,L}$};
				
		\draw[rounded corners=10pt, dashed, fill=blue!5] (7.1,5.2) rectangle (10.5,4.1);
		\draw (7.5,4.6) node{\footnotesize $u_1$ };
		\draw[dashed,->] (9.1,4.7)--(8.2,4.7);
		\draw (9.8,4.6) node{\footnotesize  $y_{in}$ };

		\draw[rounded corners=10pt,dashed, fill=blue!5] (2.5,1.4+3.8)--(6,1.4+3.8)--(8.3,-.6+3.8+.2)--(7.7,-1.4+3.8+.1)--(5.8,.3+3.8)--(2.5,.3+3.8)--cycle;
		\draw (3.2,.8+3.8) node{\footnotesize $\lambda_{1}$ };
		\draw[dashed,->] (4,.8+3.8)--(4.7,.8+3.8);
		\draw (5.5,.8+3.8) node{\footnotesize $w_{1}$ };
		\draw[dashed,->] (6.2,.5+3.85)--(7.1,-.25+3.85);
		\draw (7.5,-.7+3.8+.1) node{\footnotesize $u_2$};

		\draw[rounded corners=10pt,dashed, fill=blue!5] (2.5,1.4+2.3)--(6,1.4+2.3)--(8.3,-.6+2.3)--(7.5,-1.4+2.3)--(5.8,.3+2.3)--(2.5,.3+2.3)--cycle;
		\draw (3.2,.8+2.3) node{\footnotesize $\lambda_{2}$ };
		\draw[dashed,->] (4,.8+2.3)--(4.7,.8+2.3);
		\draw (5.5,.8+2.3) node{\footnotesize $w_{2}$ };
		\draw[dashed,->] (6.2,.5+2.3)--(7.1,-.25+2.3);
		\draw (7.5,-.7+2.3) node{\footnotesize $u_3$};

		\draw (3.2,2.1) node{\footnotesize$\vdots$};
		\draw (5.5,2.1) node{\footnotesize$\vdots$};
		\draw (7.5,0.8) node{\footnotesize$\vdots$};

		\draw[rounded corners=10pt,dashed, fill=blue!5] (2.5,1.4)--(6,1.4)--(8.3,-.6)--(7.5,-1.4)--(5.8,.3)--(2.5,.3)--cycle;
		\draw (3.2,.8) node{\footnotesize $\lambda_{L-1}$ };
		\draw[dashed,->] (4,.8)--(4.7,.8);
		\draw (5.5,.8) node{\footnotesize $w_{L-1}$ };
		\draw[dashed,->] (6.2,.5)--(7.1,-.25);
		\draw (7.5,-.7) node{\footnotesize $u_L$};

		\draw[rounded corners=10pt, dashed, fill=blue!5] (2.5,-1.2) rectangle (6.,-.1);
			\draw (3.2,-.7) node{\footnotesize $\lambda_{L}$ };
		\draw[dashed,->] (4.,-.7)--(4.7,-.7);
		\draw (5.5,-.7) node{\footnotesize  $w_{L}$ };

		\draw (-1,2) node{\small $\Lambda_L$};
		\draw[->,dashed] (-0.5,2.) --(2.3,4.6);
		\draw[->,dashed] (-0.5,2.) --(2.3,3.1);
		\draw[->,dashed] (-0.5,2.0) --(2.3,0.8);
		\draw[->,dashed] (-0.5,2.) --(2.3,-.55);
		
		\draw[very thick,->] (8.5,3.1)--(9.2,3.1);
		\draw[very thick,->] (8.5,1.6)--(9.2,1.6);
		\draw[very thick,->] (8.5,-.7)--(9.2,-.7);
		\draw[very thick,->] (4.5,-1.2)--(4.5,-1.5)--(5.7,-1.5);
	\end{tikzpicture}

	\caption{Data dependency and processor assignment for the parallel computation of
 $\mathcal{M}\cdot \Lambda_L$. Note that $u_1$ is independent
 of $\Lambda_L$ and is calculated only once when forming the
 right-hand side ${\bf b}$ in \eqref{eq:linearsystem1}.
 }
	\label{fig:processors}
\end{figure}

For a positive integer $L$, we consider the non-overlapping sub-intervals $\left(T_{\ell-1},T_{\ell}\right), \ell=1,\dots,L$ of $(0,T)$ with $T_{\ell}=\ell\Delta T$ and $\Delta T=T/L$. We define two sets of intermediate states $(Y_{\ell})_{\ell=1,\dots, L}$ and $(\Lambda_{\ell})_{\ell=1,\dots,L}$, corresponding to 
the state $y$ and the adjoint $\lambda$ at times  $T_{1},\dots,T_L$ . 
We now introduce a new parallelization idea for solving \eqref{eq:systemoptimality} inspired by the ParaExp algorithm \cite{doi:10.1137/110856137}, namely to decompose the ODEs into homogeneous and inhomogeneous parts. 
First, we consider the homogeneous sub-problems on $\lambda$, which require the backward propagation of $\lambda$ over each $(T_{\ell-1},T_L)$:
\begin{equation}
	\dot\lambda_{\ell}(t)=-{\cal L}^T\lambda_{\ell}(t), \;\; \lambda_{\ell}(T_L)=\Lambda_L, \;\;\textrm{on} \; (T_{\ell-1}, T_L),
	\label{eq:homogeneoussubproblem2}
	\end{equation}
 Next, we define the inhomogeneous part of the problem on $y$, given by 
\begin{equation}
	\dot w_{\ell}(t)={\cal L}w_{\ell}(t)-\frac{1}{\alpha}\lambda_{\ell}(t),\;\; w_{\ell}(T_{\ell-1})=0,\;\textrm{on}\;\,(T_{\ell-1},T_{\ell}), \label{eq:inhomogeneoussubproblem}
\end{equation}
for $\ell=1,\dots, L$. Note that the $w_\ell$ are solved with no initial conditions; they are instead carried by the homogeneous sub-problems $u_{\ell}$ defined below, which perform the forward propagation of $y$ over $(T_{\ell-1},T_{\ell})$:
\begin{equation}\label{eq:homogeneoussubproblem1a}\begin{split}
	\dot u_1(t)=&{\cal L}u_1(t), \; u_1(T_0)=y_{in},\;\textrm{on}\;  (T_0,T_L) \\
	\dot u_{\ell}(t)=&{\cal L}u_{\ell}(t), \; u_{\ell}(T_{\ell-1})=w_{\ell-1}(T_{\ell-1}),\;\textrm{on}\;(T_{\ell-1}, T_L),
	\end{split}	
	\end{equation}
	for $\ell=2,\dots, L$.
Finally, $y(T_\ell)$ 
is obtained by superposition:
\begin{equation}\label{exp:statesupproblem}
y(T_\ell)=w_{\ell}(T_\ell)+\sum_{j=1}^{\ell}u_j(T_\ell)
,\qquad \ell=1,\dots, L.
\end{equation}
Figure \ref{fig:processors} summarizes the dependency between $\lambda_\ell$, $w_\ell$ and $u_\ell$.
Thus, for $L$ available processors, it is natural to assign $\lambda_{\ell}, w_\ell$ and $u_{\ell+1}$ to the same processor for $\ell=1,\dots,L-1$ and $\lambda_L, w_L$ and $u_1$ to the $L^{th}$ processor. 
Note that to calculate $y(T_\ell)$, the processors only need to
exchange the values of $u$ and $w$
at the interfaces $T_\ell$, rather than whole trajectories.

Once the $Y_\ell = y(T_\ell)$ and $\Lambda_{\ell+1}=\lambda(T_{\ell+1})$ are found, the trajectories $y(t)$ and $\lambda(t)$ on $t\in [T_\ell,T_{\ell+1}]$ can easily be constructed in parallel using \eqref{eq:systemoptimality}.

Let ${\cal P}_\ell$ and ${\cal Q}_\ell$ be the solution operators for $u_\ell$ and $\lambda_\ell$ defined by the homogeneous sub-problems \eqref{eq:homogeneoussubproblem1a} and \eqref{eq:homogeneoussubproblem2}. (Even though ${\cal L}$ is identical on each interval, we write the index $\ell$ explicitly to indicate the sub-interval on which the propagation is performed.) In other words, the solution of 
\eqref{eq:homogeneoussubproblem2} is given by
\begin{equation}\label{exp:homogeneoussubproblemb}
\lambda_\ell(t)={\cal Q}_\ell(t)\cdot \Lambda_L, \;\, \textrm{on}\;\,(T_{\ell-1}, T_L),
\end{equation}
and the one of \eqref{eq:homogeneoussubproblem1a} is given by
\begin{equation}\label{eq:homogeneoussubproblema}
\begin{split}
    u_1(t)=&{\cal P}_1(t)\cdot y_{in},\;\; \textrm{on}\quad (T_0,T_L)\\
    u_{\ell}(t)=&{\cal P}_{\ell}(t)\cdot w_{\ell-1}(T_{\ell-1}),\;\;\textrm{on}\quad (T_{\ell-1},T_L).
    \end{split}
\end{equation}
Thus, substituting \eqref{exp:homogeneoussubproblemb} into \eqref{eq:inhomogeneoussubproblem} on $(T_{\ell-1},T_{\ell})$,  we get
\begin{equation}\label{eq:inhomogeneousproblem2}
\dot w_{\ell}(t)={\cal L}w_{\ell}(t)-\frac1\alpha{\cal Q}_{\ell}(t)\cdot\Lambda_{L},\quad w_{\ell}(T_{\ell-1})=0.
\end{equation}

Since $w_{\ell}(T_{\ell-1})=0$, the above implies
\begin{equation*}
w_{\ell}(t)=-\frac1\alpha\left[\int_{T_{\ell-1}}^te^{(t-s){\cal L}}{\cal Q}_{\ell}(s)ds\right]\cdot\Lambda_L.
\end{equation*}
Let  ${\cal R}_{\ell}$ denote the solution operator of $w_{\ell}$ on $(T_{\ell-1},T_{\ell})$, such that \begin{equation}\label{exp:inhomogeneoussubproblem}
w_{\ell}(t)=-\frac{1}{\alpha}{\cal R}_{\ell}(t)\cdot \Lambda_L.\end{equation} 
Therefore, substituting \eqref{exp:inhomogeneoussubproblem} into \eqref{eq:homogeneoussubproblema}, we obtain
\begin{equation}\label{exp:homogeneoussubproblema}
\begin{split}
u_1(t)=&{\cal P}_{1}(t)\cdot y_{in}\;\, \textrm{on}\;\, (T_0,T_L),\\
u_{\ell}(t)=&-\frac{1}{\alpha}{\cal P}_{\ell}(t)\cdot {\cal R}_{\ell-1}(T_{\ell-1})\cdot\Lambda_L,\;\, \textrm{on}\;\; (T_{\ell-1},T_L),
\end{split}
\end{equation} 
and inserting \eqref{exp:inhomogeneoussubproblem} and \eqref{exp:homogeneoussubproblema} into  \eqref{exp:statesupproblem} yields
\begin{equation*}
\begin{split}
y(T_{\ell})=&{\cal P}_1(T_{\ell})\cdot y_{in}-\frac1\alpha{\cal R}_{\ell}(T_{\ell})\cdot \Lambda_L\\
&-\frac{1}{\alpha}\sum_{j=2}^{\ell}{\cal P}_j(T_{\ell})\cdot {\cal R}_{j-1}(T_{j-1})\cdot \Lambda_L,\;\, \textrm{on}\;\; (T_{\ell-1},T_{\ell}).
\end{split}
\end{equation*}
Hence, the optimality system \eqref{eq:systemoptimality} in terms of $Y_{\ell} \,=\, y(T_{\ell})$ and $\Lambda_{\ell} \,=\, \lambda(T_{\ell})$  becomes
\begin{equation}\label{exp:stateunkwnons}
\begin{split}
Y_{\ell}=&{\cal P}_1(T_{\ell})\cdot y_{in}-\frac{1}{\alpha}{\cal R}_{\ell}(T_{\ell})\cdot\Lambda_L\\
&-\frac{1}{\alpha}\sum_{j=2}^{\ell}{\cal P}_j(T_\ell)\cdot{\cal R}_{j-1}(T_{j-1})\cdot\Lambda_L,\; \ell=1,\dots,L, 
\end{split}
\end{equation}
and
$\Lambda_{\ell}={\cal Q}_{\ell}(T_{\ell})\cdot\Lambda_L,\; \ell=1,\dots,L-1.$
Now that we expressed the $Y_{\ell}$ and $\Lambda_{\ell}$ in terms of a single unknown $\Lambda_L$, we only need
to impose one last equation to close the system: we use
the final condition in \eqref{eq:systemoptimality}, given by \begin{equation}\label{eq:initialcondition}\Lambda_L-Y_L+y_{tg}=0.\end{equation}  
Substituting \eqref{exp:stateunkwnons} for $\ell=L$ into the above, we obtain
\begin{equation}\label{eq:linearsystem1}
{\cal M}\cdot \Lambda_L+\textbf{b}=0,
\end{equation}
where $\textbf{b}:=y_{tg}-{\cal P}_1(T_L)\cdot y_{in}$ and 
\[
{\cal M}:=I+\frac{1}{\alpha}{\cal R}_{L}(T_L)+\frac{1}{\alpha}\sum_{j=2}^{L}{\cal P}_j(T_L)\cdot{\cal R}_{j-1}(T_{j-1}).
\]
with $I$ being the identity matrix. %
Just like for \eqref{exp:statesupproblem}, the matrix-vector product ${\cal M}\cdot\Lambda_L$ can be computed
fully in parallel:
each term of the form ${\cal P}_j(T_L)\cdot{\cal R}_{j-1}(T_{j-1})\cdot \Lambda_L$ in the sum above can be performed on the $(j-1)^{th}$ processor for $j=2,\dots,L$, while the term  ${\cal R}_{L}(T_L)\cdot \Lambda_L$ can be handled by the $L^{th}$ processor. 

In practice, the solution operators ${\cal P}_\ell$ and ${\cal R}_\ell$ are matrix exponentials that need
to be approximated numerically. Computing this approximation efficiently is a non-trivial task:
for small and
medium-sized dense matrices ($r\leq 1000$), various methods can be found in \cite{doi:10.1137/S00361445024180,doi:10.1137/1.9780898717778} and \cite{golub}. For large sparse matrices, however, the
task 
frequently requires an understanding of the spectral properties of the 
underlying operator ${\cal L}$. One possibility proposed by \cite{doi:10.1137/110856137} is projection-based methods. There, the authors claim that ``we clearly find that it is beneficial to use the Arnoldi method rather than a time-stepping method for the propagation of the linear homogeneous problems. Note that the difference between the projection and time-stepping methods gets large for high accuracy,'' meaning that the Arnoldi method can produce accurate approximations efficiently.

\section{Preconditioner design}
\label{preconditioner}
 It remains to solve the linear system \eqref{eq:linearsystem1} using an iterative method, e.g.~GMRES.
  Since the matrix ${\cal M}$ is generally ill-conditioned, a preconditioner is needed, which we
 now construct for the semi-discretized heat and wave equations.
\subsection{Heat equation}
\label{sec:preconditionerHeatEqcase}
In this section, we propose a preconditioner $\widehat{\cal M}^{-1}$ for the heat equation to speed up the convergence of \eqref{eq:linearsystem1} in GMRES. Our approach is inspired by the work in the Master's thesis \cite{Colin}, where the author derived a preconditioner based on the behavior of  ${\cal L}$ at low and high frequencies. Let us consider the heat equation given by 
\begin{equation}\label{eq:heatequ}
\dot{y}=\Delta y+\nu, \;\textrm{on} \;\; \Omega \times (0, T),
\end{equation}
with $y(x,0)=y_{in}(x)$ on $\Omega$. We consider the 1D case where $\Omega= (0,1)$ with  Dirichlet boundary conditions $y(0,t)=y(1,t)=0$,  $ t \in (0,T)$ with $T\geq 1$. Then, a semi-discretization in space of \eqref{eq:heatequ} using second-order centered finite differences leads to the following ODE system
\begin{equation}\label{eq:semiheat}
\dot y(t)={\cal L} y(t)+\nu(t),\; y(0)=y_{in},\; t\in (0,T),
\end{equation}
where $y(t), y_0, \nu(t) \in \R^{r}$, with $r$ being the number of unknowns in space and ${\cal L}$ being the discretization of $\Delta$ in space. Since ${\cal L} = {\cal L}^T$, the optimality 
system \eqref{eq:systemoptimality} becomes
\begin{equation}\label{eq:optSysHeat}
\dot y={\cal L} y-\frac{1}{\alpha}\lambda,\;  \; \dot\lambda=-{\cal L}\lambda, 
\end{equation}
with $y(0)=y_{in}$ and $\lambda(T)=y(T)-y_{tg}$.
 
To obtain an effective preconditioner, we need to study the eigenvalue properties of the matrix ${\cal M}$
in \eqref{eq:linearsystem1}. To do so, we consider
the eigenvalue decomposition ${\cal L}={\cal U}{\cal D}{\cal U}^T$, where ${\cal D}$ is a diagonal matrix with the eigenvalues of ${\cal L}$ and ${\cal U}$ a unitary matrix.  The transformation $y(t) \longmapsto {\cal U}^T\cdot y(t)$ and $\lambda(t)\longmapsto {\cal U}^T \cdot \lambda(t)$ allows us to diagonalize  \eqref{eq:optSysHeat} and obtain scalar equations of the form
\begin{equation}\label{eq:optSysHeat1D}
\dot y(t)=\sigma y(t)-\frac{1}{\alpha}\lambda(t),\;\; \dot \lambda(t)=-\sigma \lambda(t),\;\; t\in (0, T),
\end{equation}
with $y(t), \lambda(t) \in \R$ and $\sigma \,\,\leq 0$ an eigenvalue of ${\cal L}$
(see \cite{doi:10.1137/1.9780898717853}).
From \eqref{eq:optSysHeat1D}, it is easy to see that $\lambda(t)=e^{\sigma(T-t)}\lambda(T)$ and \[y(t)=e^{\sigma t}y_{in}-\frac{1}{\alpha}\int_{0}^{t}e^{\sigma(t-s)}\lambda(s)ds.\]
Substituting $\lambda(t)$ into $y(t)$, we obtain
 \[y(t)=e^{\sigma t}y_{in}-\frac{1}{\alpha}e^{\sigma(T+t)}\left(\int_{0}^{t}e^{-2\sigma s}ds\right)\lambda(T).\]
Evaluating $y$ at $t=T$ and inserting the result into the final condition $\lambda(T)=y(T)-y_{tg}$ of \eqref{eq:optSysHeat1D} (where $\Lambda_L=\lambda(T)$), we get 
 \begin{equation}\label{eq:fcontinuous}f(\sigma)\Lambda_L=e^{\sigma T}y_{in}-y_{tg},\end{equation}
where $f(\sigma):=1+\left(e^{2\sigma T}-1\right)/2\alpha\sigma$. 
 
Note that $f(\sigma)$ is an eigenvalue of ${\cal M}$ whenever $\sigma$ is an eigenvalue of ${\cal L}$. Thus, if $\textrm{Sp}({\cal L})=\{\sigma_j, j=1,\dots,r\}$ is the discrete approximation of the continuous eigenvalues of $\Delta$ given by $-j^2\pi^2$, then  $f(\sigma_j)\approx 1-1/{2\alpha\sigma_j}, j=1,\dots, r$, so that ${\cal U}f({\cal D}){\cal U}^T\approx (I-\frac{1}{2\alpha}{\cal L}^{-1})$. Hence, we propose the following preconditioner 
  \begin{equation}\label{exp:precond}\widehat{\cal M}^{-1}={\cal L}({\cal L}-\frac{1}{2\alpha}I)^{-1}.\end{equation} 
Note that there is no longer any mention of the eigenvalue decomposition in the definition of $\widehat{\cal M}$, meaning
that no such computation is needed when applying the preconditioner.
Each matrix-vector multiplication of the form $\widehat{\cal M}^{-1}\cdot\Lambda_L$ only requires a multiplication by ${\cal L}$ and the solution of an elliptic problem of the form $({\cal L}-\frac{1}{2\alpha}I)\boldsymbol{v}=\textbf{f}$, which
can be done cheaply using e.g.~algebraic multigrid \cite{ruge1987algebraic}.
 
The following result shows that for small $\delta t$, the eigenvalues of the preconditioned matrix is clustered
around 1. Thus, preconditioned GMRES is expected to converge in a small number of iterations \cite{benzi2002preconditioning}.
\begin{theorem}\label{theo:cond}
	Let a positive integer $N$ be given and ${\cal R}_{\ell}$ be approximated using the Implicit Euler method with $N$ fine sub-intervals over each $(T_{\ell-1}, T_{\ell})$.  Then any eigenvalue $\mu$ of ${\cal M}\widehat{\cal M}^{-1}$ satisfies
	\begin{equation}\label{eq:estimate}
	1<\mu <1+\frac{\delta t}{\alpha},
	\end{equation}
	where $\delta t=T/LN$.
\end{theorem}
\begin{proof} 
	By definition, ${\cal R}_{\ell}(T_{\ell})$ for \eqref{eq:optSysHeat1D} is given by 
	\[{\cal R}_{\ell}(T_{\ell})=e^{(T+T_{\ell})\sigma}\int_{T_{\ell-1}}^{T_{\ell}}e^{-2\sigma s}ds.\]
	Then, the discrete form of ${\cal R}_{\ell}(T_{\ell})$ using Implicit Euler reads 
	${\cal R}_{\ell}(T_{\ell})=\delta t e^{(T+T_{\ell}-2T_{\ell-1})\sigma}\sum_{n=1}^{N}e^{-2n\sigma\delta t},$ so that,
	\begin{equation}\label{exp:Rl}
	{\cal R}_{\ell}(T_{\ell})=\delta t e^{2(\Delta T-\delta t)\sigma}e^{(L-\ell)\Delta T\sigma}\left(\frac{e^{-2\Delta T\sigma}-1}{e^{-2\delta t\sigma}-1}\right).
	\end{equation}
	Next, from \eqref{exp:stateunkwnons}, we have
	\begin{equation}\label{exp:YL}
	\begin{split}
	Y_{L}=&{\cal P}_1(T)\cdot y_{in}-\frac{1}{\alpha}{\cal R}_{L}(T)\Lambda_L\\
	&\phantom{000000000}-\frac{1}{\alpha}\sum_{j=2}^{L}{\cal P}_j(T){\cal R}_{j-1}(T_{j-1})\Lambda_L.
	\end{split}
	\end{equation}
	Since ${\cal P}_{\ell}(T)=e^{(T-T_{\ell-1})\sigma}=e^{(L-\ell+1)\Delta T \sigma}$, replacing \eqref{exp:Rl} into \eqref{exp:YL}, we obtain 
	\begin{align*}
	Y_L=&e^{L\Delta T\sigma}\cdot y_{in}\\
	&\phantom{}-\frac{\delta t}{\alpha}e^{2(\Delta T-\delta t)\sigma}\left(\frac{e^{-2\Delta T\sigma}-1}{e^{-2\delta t\sigma}-1}\right)\left(\frac{e^{2T\sigma}-1}{e^{2\Delta T\sigma}-1}\right)\Lambda_L\\
	=&e^{L\Delta T\sigma}\cdot y_{in}-\frac{\delta t}{\alpha}\left(\frac{1-e^{2T\sigma}}{1-e^{2\delta t\sigma}}\right)\Lambda_L.
	\end{align*}
	Therefore, inserting $Y_L$ into $\Lambda_L=Y_L-y_{tg}$, we get	$f_{\delta t}(\sigma)\Lambda_L=e^{T\sigma}y_{in}-y_{tg},$
	where $f_{\delta t}(\sigma):=1+\frac{\delta t}{\alpha}\left(\frac{1-e^{2T\sigma}}{1-e^{2\delta t\sigma}}\right)$. Note that for $\sigma \in \textrm{Sp}({\cal L})$, $f_{\delta t}(\sigma)$  is eigenvalue of the discrete ${\cal M}$.
	
	The largest eigenvalue of ${\cal L}$ as function of $r$ is given by $\sigma_1(r)=-4(r+1)^2\sin^2(\frac{\pi}{2(r+1)}), r\geq 1$ (see \cite{doi:10.1137/1.9780898717853}). Then, \begin{align*}\sigma'_{1}(r)=&-8(r+1)\sin\left(\frac{\pi}{2(r+1)}\right)\\
	&\times \left[\sin\left(\frac{\pi}{2(r+1)}\right)-\frac{\pi}{2(r+1)}\cos\left(\frac{\pi}{2(r+1)}\right)\right].\end{align*}
	
	Since $r\geq 1$, $\frac{\pi}{2(r+1)} \in (0,\frac{\pi}{4}]$. Using the fact that $\tan\theta -\theta>0$, i.e., $\sin\theta-\theta\cos\theta>0$ for $0\leq \theta\leq \frac{\pi}{4}$, we conclude that $\sigma'_1(r)<0$. Thus, $\sigma_1(r)$ decreases on $(1, +\infty)$ and its maximum is $\sigma_1(1)=-8$. 
	
	Now, let $\psi_0(\sigma):= f_{\delta t}(\sigma)f^{-1}(\sigma)$ for $\sigma \leq -8$. Then, 
	$\psi_0(x)-1={\varphi(\sigma)}/{(2\alpha\sigma-1)(1-e^{2\delta t\sigma})},
	$
	where $\varphi(\sigma): =2\sigma\delta t(1-e^{2T\sigma})+(1-e^{2\delta t\sigma})$. We have $\varphi'(\sigma)=2\delta t(1-e^{2\delta \sigma})-2\delta t(2\sigma T+1)e^{2T\sigma}>0,$ since $(2\sigma T+1)<0$. But $\varphi(-8)=-16\delta t(1-e^{-16T})+(1-e^{-16\delta t})<0$ for $ \delta t>0$, which means $\varphi(\sigma)<0$
    for $\sigma \leq -8$. Thus, $\psi_0(\sigma)-1>0,$ i.e., $\psi_0(\sigma)>1$.
	
Next, we consider $f^0_{\delta t}(\sigma):=1+\frac{\delta t}{\alpha\left(1-e^{2\delta t\sigma}\right)}\geq f_{\delta t}(\sigma)$ 
	 and  $\psi(\sigma):=f^0_{\delta t}(\sigma)f^{-1}(\sigma)$ for $\sigma \in (-\infty,0)$, whose derivative is
  $\psi'(\sigma)={\psi_1(\sigma)}/{\alpha(2\delta t\alpha-1)^2(1-e^{2\delta t \sigma})^2},$
 where $\psi_1(\sigma):=-2\alpha(1-e^{2\delta t \sigma})^2-\delta t(1-e^{2\delta t\sigma})+4\delta t^2 \sigma(2\alpha \sigma-1)e^{2\delta t \sigma}.$
 We have $\psi_1'(\sigma)=8\delta t e^{2\delta t \sigma}\psi_2(\sigma)$, where $\psi_2(\sigma):=\mbox{$\alpha(1-e^{2\delta t\sigma})$}+\sigma\delta t^2(2\alpha \sigma-1)+2\alpha\delta t\sigma$. Since $\psi_2''(\sigma)=4\alpha\delta t^2(1-e^{2\delta t\sigma})>0$ and $\psi_2'(0)=-\delta t^2<0$, we get $\psi_2'(\sigma)<0$. Also, since $\psi_2(0)=0$, $\psi_2(\sigma)\geq 0$, which implies $\psi_1'(\sigma)>0$. Finally, since $\psi_1(0)=0$, $\psi_1(\sigma)\leq 0$ and $\psi'(\sigma)\leq 0$. Hence, $\psi$ is decreases on $(-\infty,0)$ and tends to $1+\delta t/\alpha$ as $\sigma$ tends to $-\infty$, such that $\psi(\sigma)<1+\frac{\delta t}{\alpha}$.
 
 Now, the quantity $\psi_0(\sigma)$ is the eigenvalue of ${\cal M}\widehat{\cal M}^{-1}$ associated with the fixed $\sigma \in \textrm{Sp} ({\cal L})$. Then we have $1<\psi_0(\sigma)\leq \psi(\sigma)\leq 1+\frac{\delta t}{\alpha}$, which concludes the proof. 
\end{proof}

\begin{remark}\label{rem:precond}
	Note that  $\widehat{\cal M}^{-1}$ is derived from the continuous form of ${\cal M}$. Therefore, we expect the preconditioner to be more efficient for a high-order approximation of ${\cal R}_{\ell}$ than for a lower-order one.
\end{remark}

Analogously, we can prove that \eqref{exp:precond} yields an effective preconditioner for \eqref{eq:linearsystem1} when the boundary condition in \eqref{eq:heatequ} is Dirichlet-Neumann in 1D or Dirichlet in 2D.

In our numerical test, we use the example studied in \cite{paraopt}: we set $T=1$, $L=10$, $r=100$, $y_{in}(x)=\exp(-100(x-1/2)^2)$ and $y_{tg}(x)=\frac12\exp(-100(x-1/4)^2)+\frac12(-100(x-3/4)^2)$ with Dirichlet boundary conditions. We use Matlab R2021b as our numerical environment.
Moreover, to get ${\cal R}_{\ell}$, we use two quadrature formulas, namely Implicit Euler and the Singly Diagonal Implicit Runge-Kutta (SDIRK) method of order $3$, whose stages and weights are $c=[1/2+\sqrt 3/6, 1/2-\sqrt 3/6]$ and $d=[1/2,1/2]$ respectively. 

\begin{figure}[h!]
	\centering
	\includegraphics[scale=0.4]{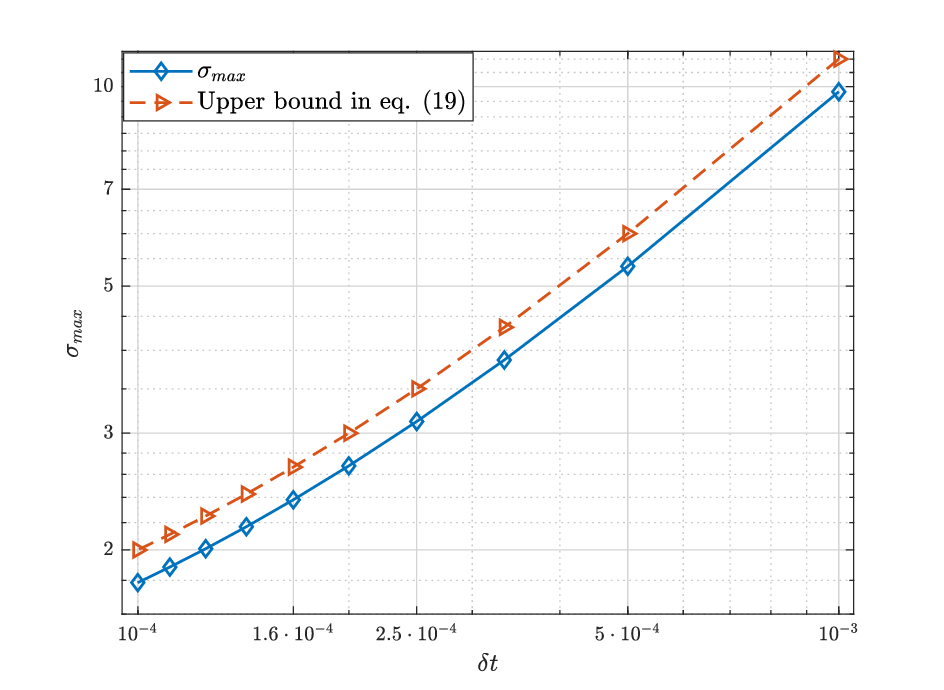}
	\caption{The maximal eigenvalue of ${\cal M}\widehat{\cal M}^{-1}$ and the upper bound in \eqref{eq:estimate} for various $\delta t$ with $\alpha=10^{-4}$.}
	\label{fig:fig1}
\end{figure}

First, we investigate the numerical behavior of the upper bound in \eqref{eq:estimate} and the maximal eigenvalue $\sigma_{max}$ of $ {\cal M}\widehat{\cal M}^{-1}$ for various $N=100k$, $k=1,\dots,10,$ i.e., $\delta t=1/N$ and fixed $\alpha=10^{-4}$ with only Implicit Euler for getting ${\cal R}_{\ell}$.
 The result is shown in Fig. \ref{fig:fig1}, where we observe that the upper bound in \eqref{eq:estimate} follows closely the evolution of ${\sigma_{max}}$ for various $\delta t$. Thus, \eqref{eq:estimate} provides an efficient estimation of the upper bound of the eigenvalues of ${\cal M}\widehat{\cal M}^{-1}$.

Next, we investigate the convergence speed of the linear system \eqref{eq:linearsystem1} in GMRES for $N=1000$ and $\alpha=\{10^{-4},10^{-6}\}$. We use for this instance the function \texttt{gmres} in Matlab with \texttt{restart=1}, \texttt{tol=1e-8} and \texttt{maxit=500}.  

For $\alpha=10^{-4}$, the result is shown in Table \ref{tab:1}. We observe that when solving \eqref{eq:linearsystem1}, unpreconditioned GMRES reaches 500 iterations without satisfying the tolerance for either quadrature formula. Moreover, the eigenvalues of ${\cal M}$ for both quadrature formulas lie in $(1,500)$. In contrast, for the preconditioned case, when SDIRK is used, GMRES reaches the tolerance with only $2$ iterations, and $9$ iterations for the Implicit Euler case.  
We observe a similar behavior for $\alpha=10^{-6}$, with poor convergence
without preconditioning, fairly good convergence for Implicit Euler (44 iterations)
and excellent convergence for SDIRK (4 iterations). The results are shown in Table \ref{tab:2}. 

\begin{table}[h!]
	\caption{GMRES for solving \eqref{eq:linearsystem1} for the heat equation, with $\alpha=10^{-4}$.}
	\label{tab:1}
	\centering
	\begin{tabular}{c}
		Matrix ${\cal M}$\\
	
\begin{tabular}{lcccc}
	\hline\hline
	& $\sigma_{min}$&$\sigma_{max}$ &\# Iters &Res\\
	\hline	
	Euler& 2.0&4.9e2&500 &1.3e-8\\
	SDIRK&1.08&4.9e2&500&4.08e-7\\
	\hline\hline
\end{tabular}\\
\quad\\
	Preconditioned matrix ${\cal M}\widehat{\cal M}^{-1}$\\
	
	\begin{tabular}{lcccc}
		\hline\hline
		& $\sigma_{min}$&$\sigma_{max}$ &\# Iters &Res\\
		\hline
		Euler& 1.0&1.78&9 &7.7e-9\\
		SDIRK&0.97&1.0&2&9.64e-9\\
		\hline\hline
	\end{tabular}
\end{tabular}
\vspace{10pt}
		\caption{GMRES for solving \eqref{eq:linearsystem1} for the heat equation, with $\alpha=10^{-6}$.}
	\label{tab:2}
	\centering
	\begin{tabular}{c}
		Matrix ${\cal M}$\\
		
		\begin{tabular}{lcccc}
			\hline\hline
			& $\sigma_{min}$&$\sigma_{max}$ &\# Iters &Res\\
			\hline	
			Euler& 1e2&4.97e4&500 &1.08e-6\\
			SDIRK&9.68&4.96e4&500&4.2e-5\\
			\hline\hline
		\end{tabular}\\
		\quad\\
		Preconditioned matrix ${\cal M}\widehat{\cal M}^{-1}$\\
		
		\begin{tabular}{lcccc}
			\hline\hline
			& $\sigma_{min}$&$\sigma_{max}$ &\# Iters &Res\\
			\hline
			Euler& 1.0&7.76&44&8.37e-9\\
			SDIRK&0.74&1.0&4&7.31e-9\\
			\hline\hline
		\end{tabular}
	\end{tabular}
\end{table}

In both Tables \ref{tab:1} and \ref{tab:2}, we observe that preconditioned GMRES converges more quickly for SDIRK than for Implicit Euler. This is consistent with Remark \ref{rem:precond}, since SDIRK is of order $3$, which is higher than the order $1$ of Implicit Euler. 

Theorem \ref{theo:cond} asserts that the condition number $cond({\cal M}\widehat{\cal M}^{-1})$ is bounded by $1+\frac{\delta t}{\alpha}$ when Implicit Euler is used to compute ${\cal R}_\ell$. 
Since ${\cal M}\widehat{\cal M}^{-1}$ is symmetric (as ${\cal M}$ and 
$\widehat{\cal M}$ commute and are both symmetric), we deduce that the number of preconditioned GMRES iterations remains bounded as $r\to \infty$, which is confirmed by the results in  Table \ref{tab:10}. We also observe that 
GMRES converges faster for a higher order method like SDIRK than for Implicit Euler. 

 \begin{table}[h!]
 	\centering
	\caption{GMRES for solving \eqref{eq:linearsystem1} with preconditioning for the heat equation, with  $\alpha=10^{-4}$ and for various $r$.}
	\label{tab:10}
		\begin{tabular}{l|cc|cc}
			\hline\hline
			\multirow{2}{*}{$r$}& \multicolumn{2}{|c}{\# Iters(Euler)}&\multicolumn{2}{|c}{\# Iters(SDIRK)}\\
   \cline{2-5}
                &$L=10^3$&$L=3\cdot10^3$&$L=10^3$&$L=3\cdot10^3$\\
			\hline
			100& 9&6&3&2\\
                200&10&7&3&3\\
                250&11&7&3&3\\
                600&11&7&3&3\\
			\hline\hline
		\end{tabular}
\end{table}
 Finally, one can use an explicit method as a numerical integrator to approximate ${\cal R}_{\ell}$,  as long
 as a CFL condition is satisfied.

\subsection{Wave equation}
\label{sec:preconditionerWaveEqcase}
We now consider the optimal control problem constrained by the 1D wave equation given by 
\begin{equation}\label{eq:wave}
   \partial_{tt}u=\Delta u+\nu,\; \textrm{on}\;\, (0,1) \times (0,T),
\end{equation} 
with $u(x,0)=u_0(x), \;\partial_tu(x,0)=0,\; x \in (0,1)$. For simplicity, we will consider homogeneous Dirichlet boundary conditions in space. Discretizing in space using second-order centered finite differences leads to the ODE system
\begin{equation}\label{eq:waveeq} \dot y
= {\cal L} y + {\cal B} \nu,\;\, \textrm{with}\;\; {\cal L} = \begin{bmatrix}0 & I\\ \Delta_h & 0\end{bmatrix},\; \; {\cal B} = \begin{bmatrix}0 \\  I\end{bmatrix},
\end{equation}
where $y= \begin{bmatrix} u_h\\ \partial_tu_h\end{bmatrix}$ and $\Delta_h$ is the discrete Laplacian.

To derive a preconditioner for the system \eqref{eq:linearsystem1}, we again
look to the continuous problem. Since ${\cal L}$ is no longer symmetric,
the continuous analogue of ${\cal M}$ takes the form
$$ {\cal M}_c := I + \frac{1}{\alpha}\int_0^Te^{s{\cal L}}{\cal B}{\cal B}^Te^{s{\cal L}^T}\,ds. $$
Let us evaluate this integral. First note that
$$ e^{s{\cal L}} = I + s{\cal L} + \frac{s^2}{2!} {\cal L}^2 + \frac{s^3}{3!} {\cal L}^3 + \cdots. $$
\newcommand{\cA}{{\cal A}}%
\newcommand{\cL}{{\cal L}}%
\newcommand{\sinc}{\mathrm{sinc}}%
Introducing the notation $A: = -\Delta_h$, we see that
$$ {\cal L}^2 = \begin{bmatrix}\Delta_h & 0\\ 0 & \Delta_h\end{bmatrix} = -
\begin{bmatrix}A & 0\\ 0 & A\end{bmatrix} =: -\cA$$
is a symmetric negative definite matrix, so we can write
\begin{align*}
e^{s{\cal L}} &= \left(I - \frac{s^2}{2!}\cA + \frac{s^4}{4!}\cA^2 +\cdots \right)\\
&\phantom{000000}+ s{\cal L}\left(I - \frac{s^2}{3!}\cA + \frac{s^4}{5!}\cA^2 +\cdots \right)\\
&= \cos(s\cA^{1/2}) + s{\cal L}\mathrm{sinc}(s\cA^{1/2}),
\end{align*}
where $\mathrm{sinc}(x) = \sin(x)/x$ is the sinc function and $\cA^{1/2}$ is the
symmetric square root of $\cA$ (i.e., a symmetric matrix with positive eigenvalues whose
square is $\cA$). This leads to
\begin{align*}
 e^{s\cL}{\cal B}{\cal B}^Te^{s\cL^T} &= [\cos(s\cA^{1/2}) + s{\cal L}\mathrm{sinc}(s\cA^{1/2})]\cdot {\cal B}\\
 &\qquad\cdot{\cal B}^T\cdot [\cos(s\cA^{1/2})+ s\;\mathrm{sinc}(s\cA^{1/2}){\cal L}^T].
\end{align*}
By direct calculation and using $x\cdot\sinc(xA^{1/2})=A^{-1/2}\sin(xA^{1/2})$, we obtain 
$$ e^{s\cL}{\cal B}{\cal B}^Te^{s\cL^T} = \begin{bmatrix} {\cal C}_{11}(s) & {\cal C}_{12}(s)\\
{\cal C}_{21}(s) & {\cal C}_{22}(s)\end{bmatrix}, $$
where ${\cal C}_{11}(s) := A^{-1}\sin^2(sA^{1/2}),\; {\cal C}_{22}(s): = \cos^2(sA^{1/2}),$ 
${\cal C}_{12}(s) := A^{-1/2}\cos(sA^{1/2})\sin(sA^{1/2})$ and ${\cal C}_{21}(s) := {\cal C}_{12}(s).$
We can now integrate these coefficients to obtain
$$
{\cal M}_c = 
\begin{bmatrix} {\cal M}_{11} & {\cal M}_{12}\\ {\cal M}_{21} & {\cal M}_{22}
\end{bmatrix},
$$
where $ {\cal M}_{12} := \frac{1}{2\alpha}A^{-1}(I-\cos(2TA^{1/2})), \; {\cal M}_{21} := {\cal M}_{12}, $
\begin{align*}
{\cal M}_{11} &:= I + \frac{T}{2\alpha}A^{-1} - \frac{1}{4\alpha}A^{-3/2}\sin(2TA^{1/2}),\\
{\cal M}_{22} &:= (1+\frac{T}{2\alpha})I + \frac{1}{4\alpha}A^{-1/2}\sin(2TA^{1/2}).
\end{align*}
To find a good preconditioner for ${\cal M}_c$, let us analyze ${\cal M}_c^{-1}$.
Since all ${\cal M}_{ij}$ are square and they all commute, we have
$${\cal M}_c^{-1} = \begin{bmatrix} {\cal H}^{-1}{\cal M}_{22} & -{\cal H}^{-1}{\cal M}_{12}\\
-{\cal H}^{-1}{\cal M}_{21} & {\cal H}^{-1}{\cal M}_{11}\end{bmatrix}, $$
with ${\cal H}: = {\cal M}_{11}{\cal M}_{22} - {\cal M}_{12}{\cal M}_{21}$. 
Just like for the heat equation, we replace $A$ by a scalar $\sigma > 0$ and study the eigenvalues of
${\cal M}_c^{-1}$ when
$\sigma \approx k^2\pi^2$ approximates the eigenvalues of the negative Laplacian. We observe for large $k$ that 
${\cal H} \approx \frac{1}{4\alpha^2}(2\alpha+T)A^{-1}\left(T I+2\alpha A\right),\;\; {\cal H}^{-1}{\cal M}_{12}={\cal H}^{-1}{\cal M}_{21}\approx 0$ and
\begin{itemize} 
\item[$\bullet$] $ {\cal H}^{-1}{\cal M}_{11}\approx I-\left(TI+2\alpha A \right)^{-1},$
\item[$\bullet$] ${\cal H}^{-1}{\cal M}_{22} \approx \left(1-\frac{1}{(2\alpha+T)}\right)I.$
\end{itemize}
These observations lead us to approximate ${\cal M}_c^{-1}$ by a block diagonal
matrix $\widehat{\cal M}^{-1}$ of the form
\begin{equation}\label{exp:precondwave}\widehat{\cal M}^{-1} :=  I-
\begin{bmatrix} (aI + bA)^{-1} & 0\\
0 & cI\end{bmatrix},\end{equation}
and use it as a preconditioner for \eqref{eq:linearsystem1}, where $a=T, b=2\alpha $ and finally $c=1/(2\alpha+T)$. 
As before, one need not compute the eigenvalue decomposition in order to
apply the preconditioner $\widehat{\cal M}$: each application only requires solving an elliptic problem of the form
$(aI+bA)\textbf{v} = \textbf{f}$, which again can be done cheaply using e.g.~multigrid.


In our numerical experiment, we take $T=2.3$, $u_0(x)=\exp(-100(x-1/2)^2)$, so that $y_{in}=(u_0(x),0)$ and $y_{tg}(x)=\left[(\frac12\exp(-100(x-1/4)^2)+\frac12(-100(x-3/4)^2),\;0\right]$. We set $L=10, r=100$, and we use the same Matlab functions for computing ${\cal R}_{\ell}$ and ${\cal P}_{\ell}$ as in Section \ref{sec:preconditionerHeatEqcase}. We first investigate the convergence speed of the linear system \eqref{eq:linearsystem1} in GMRES for $N=1000$ and $\alpha=10^{-6}$. We use for this instance the function \texttt{gmres} in Matlab with \texttt{restart=[]}, \texttt{tol=1e-8} and \texttt{maxit=size(${\cal M}$,1)}.  The results are shown in Table \ref{tab:3}, where we observe that 
unpreconditioned GMRES applied to \eqref{eq:linearsystem1} converges in $84$ iterations for both quadrature formulas, whereas the preconditioned iteration converges in only 4 iterations.
\begin{table}[h]
	\caption{GMRES for solving \eqref{eq:linearsystem1} for the wave equation, with $\alpha =10^{-6}$.}
	\label{tab:3}
	\centering
	\begin{tabular}{c}
		Matrix ${\cal M}$\\
\begin{tabular}{lccc}
    \hline\hline
	& $\textrm{cond}({\cal M})$&\# Iters &Res\\
	\hline	
	Euler &3.8e4&84 & 8.74e-9\\
	SDIRK&3.8e4&84 & 8.74e-9\\
	\hline\hline
\end{tabular}\\
	Preconditioned matrix ${\cal M}\widehat{\cal M}^{-1}$\\
	\begin{tabular}{lcccc}
		\hline\hline
		& \textrm{cond}(${\cal M}\widehat{\cal M}^{-1}$) &\# Iters &Res\\
		\hline
		Euler& 2.59&4 & 1.58e-9\\
		SDIRK& 2.59&4& 1.58e-9\\
		\hline\hline
	\end{tabular}
\end{tabular}
\end{table}

 Unlike for the heat equation, the higher order of SDIRK does not lead to better convergence compared to Implicit Euler. Nonetheless, for both methods, we have observed that the number of GMRES iterations does not change as we increase the number of time steps $N$ from 100 to 1000, as long as
 preconditioning is used. 
 
 These results show that for fixed $r$, the discretization parameters in time do not change the behavior of the convergence of \eqref{eq:linearsystem1} in GMRES. We also observe that when $r$ increases, the number of iterations of the unpreconditioned system increases, whereas the number of iterations of the preconditioned system decreases. The results are shown in Tables \ref{tab:5} and \ref{tab:4}. 
  \begin{table}[h]
	\caption{GMRES for solving \eqref{eq:linearsystem1} for the wave equation with Implicit Euler with $\alpha=10^{-6}$ for various $r$.}
	\label{tab:5}
	\centering
\begin{tabular}{ccccc}
	\hline\hline
	$r$& \textrm{cond}(${\cal M}$)&\# Iters&  $\textrm{cond}({\cal M}\widehat{\cal M}^{-1})$&\# Iters \\
	\hline	
10 &4.83e2&10&2.47&5\\
    150&7.75e4&76&2.81&3\\
    350&2.48e5&104&2.49&3\\
    \hline\hline
\end{tabular}\\[10pt]
%
 %
	\caption{GMRES for solving \eqref{eq:linearsystem1}  for the wave equation with Implicit Euler for various $\alpha$.}
	\label{tab:4}
	\centering
	
\begin{tabular}{ccccc}
	\hline\hline
	$\alpha$& \textrm{cond}(${\cal M}$)&\# Iters&  $\textrm{cond}({\cal M}\widehat{\cal M}^{-1})$&\# Iters \\
	\hline	
1e-5&2.26e4&52&2.41&3\\
    1e-3&4.98e2&20&1.09&3\\
    1e-1&6.03&9&1.01&3\\
    1e1&1.05&4&1.0&2\\
	\hline\hline
\end{tabular}\\

\end{table}

\section{Ongoing work}
%
We are currently studying the behavior of our preconditioners when ${\cal M}$ is obtained from an explicit method while respecting the CFL condition, in order to better understand the impact of using explicit versus implicit methods. We are also working on more general convergence properties of the algorithm, its error analysis, and on understanding its performance compared to existing parallel-in-time algorithms.

\section{Acknowledgments}
The work described in this paper is partially supported by a grant from
the ANR/RGC joint research scheme sponsored by the Research Grants Council of the Hong
Kong Special Administrative Region, China (Project no. A-CityU203/19), by the French National
Research Agency (Project ALLOWAPP, grant ANR-19-CE46-0013-01 and Projet DEEPNUM, grant ANR-21-CE23-0017) and by Mitacs-Globalink Canada and INRIA France (grant IT34057, 2023).


\bibliographystyle{plain}
\bibliography{biblio.bib}

\begin{thebibliography}{10}

\bibitem{barker2015domain}
Andrew~T Barker and Martin Stoll.
\newblock Domain decomposition in time for {PDE}-constrained optimization.
\newblock {\em Computer Physics Communications}, 197:136--143, 2015.

\bibitem{benzi2002preconditioning}
Michele Benzi.
\newblock Preconditioning techniques for large linear systems: a survey.
\newblock {\em Journal of Computational Physics}, 182(2):418--477, 2002.

\bibitem{doi:10.1137/1.9780898717853}
Albrecht Böttcher and Sergei~M. Grudsky.
\newblock {\em Spectral Properties of Banded Toeplitz Matrices}.
\newblock Society for Industrial and Applied Mathematics, 2005.

\bibitem{paraopt}
M.~J. Gander, F.~Kwok, and J.~Salomon.
\newblock Para{O}pt: A parareal algorithm for optimality systems.
\newblock {\em SIAM Journal on Scientific Computing}, 42(5):A2773--A2802, 2020.

\bibitem{gander201550}
Martin~J Gander.
\newblock 50 years of time parallel time integration.
\newblock In {\em Multiple Shooting and Time Domain Decomposition Methods:
  MuS-TDD, Heidelberg, May 6-8, 2013}, pages 69--113. Springer, 2015.

\bibitem{doi:10.1137/110856137}
Martin~J. Gander and Stefan G\"{u}ttel.
\newblock Paraexp: A parallel integrator for linear initial-value problems.
\newblock {\em SIAM Journal on Scientific Computing}, 35(2):C123--C142, 2013.

\bibitem{gander2007optimized}
Martin~J Gander and Laurence Halpern.
\newblock Optimized schwarz waveform relaxation methods for advection reaction
  diffusion problems.
\newblock {\em SIAM Journal on Numerical Analysis}, 45(2):666--697, 2007.

\bibitem{gander2016schwarz}
Martin~J Gander and Felix Kwok.
\newblock Schwarz methods for the time-parallel solution of parabolic control
  problems.
\newblock In {\em Domain decomposition methods in science and engineering
  XXII}, pages 207--216. Springer, 2016.

\bibitem{golub}
G.H. Golub and C.F. Van~Loan.
\newblock {\em Matrix computations}.
\newblock The John Hopkins University Press, Baltimore, 2013.

\bibitem{heinkenschloss2005time}
Matthias Heinkenschloss.
\newblock A time-domain decomposition iterative method for the solution of
  distributed linear quadratic optimal control problems.
\newblock {\em Journal of Computational and Applied Mathematics},
  173(1):169--198, 2005.

\bibitem{doi:10.1137/1.9780898717778}
Nicholas~J. Higham.
\newblock {\em Functions of Matrices}.
\newblock Society for Industrial and Applied Mathematics, 2008.

\bibitem{hoang2013space}
Thi-Thao-Phuong Hoang, J{\'e}r{\^o}me Jaffr{\'e}, Caroline Japhet, Michel Kern,
  and Jean~E Roberts.
\newblock Space-time domain decomposition methods for diffusion problems in
  mixed formulations.
\newblock {\em SIAM Journal on Numerical Analysis}, 51(6):3532--3559, 2013.

\bibitem{li2015convergence}
Shishun Li and Xiao-Chuan Cai.
\newblock Convergence analysis of two-level space-time additive {S}chwarz
  method for parabolic equations.
\newblock {\em SIAM Journal on Numerical Analysis}, 53(6):2727--2751, 2015.

\bibitem{maday2007monotonic}
Yvon Maday, Julien Salomon, and Gabriel Turinici.
\newblock Monotonic parareal control for quantum systems.
\newblock {\em SIAM Journal on Numerical Analysis}, 45(6):2468--2482, 2007.

\bibitem{doi:10.1137/S00361445024180}
Cleve Moler and Charles Van~Loan.
\newblock Nineteen dubious ways to compute the exponential of a matrix,
  twenty-five years later.
\newblock {\em SIAM Review}, 45(1):3--49, 2003.

\bibitem{ong2020applications}
Benjamin~W Ong and Jacob~B Schroder.
\newblock Applications of time parallelization.
\newblock {\em Computing and Visualization in Science}, 23:1--15, 2020.

\bibitem{ruge1987algebraic}
John~W Ruge and Klaus St{\"u}ben.
\newblock Algebraic multigrid.
\newblock In {\em Multigrid methods}, pages 73--130. SIAM, 1987.

\bibitem{toselli2005domain}
Andrea Toselli and Olof Widlund.
\newblock {\em Domain Decomposition Methods -- Algorithms and Theory}.
\newblock Springer-Verlag, Berlin, 2005.

\bibitem{Colin}
Colin~S.C. Tsang.
\newblock Preconditioners for linear parabolic optimal control problems.
\newblock Master's thesis, Hong Kong Baptist University, Hong Kong, China,
  August 2017.

\bibitem{ulbrich2007generalized}
Stefan Ulbrich.
\newblock Generalized {SQP} methods with “parareal” time-domain
  decomposition for time-dependent pde-constrained optimization.
\newblock In {\em Real-time PDE-constrained optimization}, pages 145--168.
  SIAM, 2007.

\end{thebibliography}
\end{document}